\documentclass[10pt,twocolumn,a4paper,conference]{IEEEtran}

\usepackage{cite}	
\usepackage{graphicx} 
\usepackage[latin1]{inputenc} 
\usepackage[T1]{fontenc} 
\usepackage{amsmath,amsfonts,amsbsy,amssymb} 
\usepackage{mathabx} 
\usepackage{amssymb} 
\usepackage{amsmath}
\usepackage{amsthm}	
\usepackage{mathrsfs}
\usepackage[nolist]{acronym} 
\usepackage{tabularx} 
\usepackage{multirow}
\usepackage{wasysym}
\usepackage{float}
\usepackage{color} 

\usepackage{enumitem} 

\usepackage{graphicx}
\usepackage{caption}
\usepackage{subcaption}
\usepackage{multicol}

\begin{document}

\title{Optimum Transmission Rate in Fading Channels with Markovian Sources and QoS Constraints} 
\author{ Fahad Qasmi, Mohammad Shehab, Hirley Alves, and Matti Latva-aho\\
	
	\IEEEauthorblockA{
		Centre for Wireless Communications (CWC), University of Oulu, Finland\\
	}
	Email: firstname.lastname@oulu.fi
}
%
\maketitle


\begin{abstract}
 This paper evaluates the performance of reliability and latency  in machine type communication networks, which composed of single transmitter and receiver in the presence of Rayleigh fading channel.  The source's traffic arrivals are modeled as Markovian processes namely Discrete-Time Markov process, Fluid Markov process, and Markov Modulated Poisson process,  and  delay/buffer overflow constraints are imposed. Our approach is based on the reliability and latency outage probability, where transmitter not knowing the channel condition, therefore the transmitter would be transmitting information over the fixed rate. The fixed rate transmission is modeled as a two  state Discrete time Markov process, which identifies the reliability level of wireless transmission.  Using effective bandwidth  and effective capacity theories, we evaluate the trade-off between reliability-latency and identify  QoS requirement.   
 The impact of different source traffic originated from MTC devices under QoS constraints on the effective transmission rate are investigated.     
\end{abstract}

\section{Introduction}\label{introduction}
Recent trends in research and development suggest that the Fifth Generation (5G) of mobile network may bring technology evolution in the form of expanding broadband capacity, mobility and cloud services. This evolution does not seem to be standalone, but almost every industry may need a re-definition of their business models \cite{E1}. 5G is not just extension of the 4G technologies. Therefore it not only focuses on the enhanced coverage, connectivity, data rates and spectral efficiency  but also addresses critical and massive traffic generated by machine type devices. Such devices operate with little or no human interaction and called Machine Type Communication (MTC).  
There are two operating modes proposed by 5G for MTC based applications namely massive MTC (mMTC) and ultra-reliable low latency communication (URLLC) \cite{P1}.  

MTC has attracted much interest in the recent years which promises a huge market growth with expected 50 billion connected devices by 2020 \cite{2020}. Furthermore, cellular networks are becoming a more appropriate candidate in order to fulfill the requirements of MTC applications in terms of mobility, coverage, diversity and ease of deployment. A traffic model is a stochastic process that matches the behavior of physical quantities of measured data traffic. Current cellular networks are based on standard traffic model which is designed and optimized for typical behavior of human subscribers. It is worth noting that the traffic behavior of MTC is quite different from the HTC (Human type communication) \cite{mtraffic-issue}. For example:
\begin{itemize}
	\item 	MTC Traffic is mostly uplink dominant.
	\item	MTC traffic generation frequency is typically all around the day (i.e, 24 hours) while HTC traffic flow duration is mostly during day or evening time but not at night.
	\item	MTC QoS requirement is different from HTC (i.e. different reliability and latency requirements).
	\item		MTC traffic is bursty (suddenly the volume of data flow increase in response to trigger of certain events).
	\item	MTC uses short as well as small number of packets.
%
		%
\end{itemize}
New traffic models are needed to capture the behavior of massive MTC traffic and the imposed QoS guarantees. The traffic models are mainly classified into source and aggregated traffic models. It can be said that the source traffic model is usually designed for human generated traffic that is based on Poisson distribution of arrival rate. However, it is not feasible to model the traffic generated by large amount of autonomous machines simultaneously due to the heterogeneous and uncoordinated nature of the traffic. Moreover, Poisson distribution usually fails to capture the burstiness and multimodality of the real traffic sequence. Aggregated traffic models are suitable for MTC network as they capture the traffic properties of a group of users or networks which have homogeneous and coordinated traffic characteristics \cite{Mtraffic12}.

In many MTC use cases, QoS is affected by the wireless channel change due to environment and multipath fading. Hence, source characteristics and reliable transmission rate are also time varying. To account for the time variation of service process in the queuing system, we resort to the theory of effective capacity, which is a cross layer model that can ensure QoS in time varying wireless channel. Effective capacity is defined as the maximum constant arrival rate that a given time varying service process can support while providing statistical QoS guarantees \cite{main}. Herein, we are particularly interested in using Markovain source models including discrete time Markov, Markov fluid and Markov modulated Poisson sources with effective capacity to conduct throughput analysis of random and bursty source traffic pattern in MTC framework.

Recently effective capacity of wireless communication has attracted much attention to estimate reliability, latency, security, energy efficiency and transmit power. For instance, in \cite{main2}, the authors considered fixed rate transmission technique and applied effective capacity to evaluate energy efficiency under QoS constraints. The authors used fixed-rate transmission modeled as a two-state (ON/OFF) discrete-time Markov chain. In \cite{main5}, the authors considered fixed-rate transmission modeled as a two-state (ON/OFF) continuous-time Markov chain and utilize effective capacity to analyse energy efficiency with Markov arrival under QoS constraint. In \cite{r7}, the authors considered effective capacity when the transmitter and receiver know the instantaneous channel gain. They derived the optimal power and rate adaptation technique which maximize the throughput under QoS constraint.

In this paper, we evaluate the performance of MTC in a point-to-point transmission. Using effective capacity, we design the reliability and latency aware wireless communication link layer model. We incorporate Markovian source arrival processes with reliable wireless communication model as well. Different from \cite{main5} and \cite{main3}, we use fixed-rate transmission, which is modeled as a two-state (ON/OFF) discrete-time Markov chain and effective capacity to analyze reliability and latency with Markovian arrival process under QoS constraint.  We build our contribution upon the reliability and latency framework proposed in \cite{main2}. In \cite{main2}, the authors evaluated the impact of the constant source arrival traffic on effective energy efficiency. This framework helps us to determine the level of reliability and latency for each transmission rate. We also extend the work in \cite{main} to incorporate different arrival source traffic models and estimate the reliability of the network. We consider ON and OFF states channel models to investigate the optimum transmission rate of random and bursty sources in order to maximize the system throughput.

\section{Preliminaries}
\subsection{System Model}
In the proposed work, we consider single transmitter and receiver with point-to-point link through Rayleigh block fading channel where fading coefficients vary independently from one frame to another. The discrete time input output relation of the $i^{th}$ symbol is \vspace{-1mm} 
\begin{align}\label{eq1}
y_i=h_ix_i+n_i\quad i=1,2, \ldots ,
\end{align}
where $x_i$ and $y_i$ are the channel input and output respectively. $n_i$ is zero mean, circularly symmetric, complex Gaussian random noise and $\gamma$ is the average transmitted signal-to-noise ratio. Finally, $h_i$ is the channel fading coefficient which is stationary and ergodic discrete time process. $z_i=|h_i|^2$ is the squared envelope  of Rayleigh fading block coefficients. The instantaneous channel capacity it given by
\begin{align}\label{eq2}
C_i=\rm log_2(1+\gamma \mathit{z_i} )\  bits/s.
\end{align}

In this paper, we assume that the receiver is able to estimate the channel coefficient $h_i$, whereas the transmitter does not know this information. Therefore the transmitter would be transmitting information over a fixed rate $r$ bits/s. When $r<C_i$ then the channel is considered in ON state and reliable communication is accomplished. While $r\geq C_i $ then the channel is considered in OFF state and we can not acquire reliable communication.
\begin{figure} [!t] 
	\centering \vspace{-2mm}
	\includegraphics[clip, trim= 1cm 12cm 0.5cm 10.5cm,width=0.77\columnwidth]{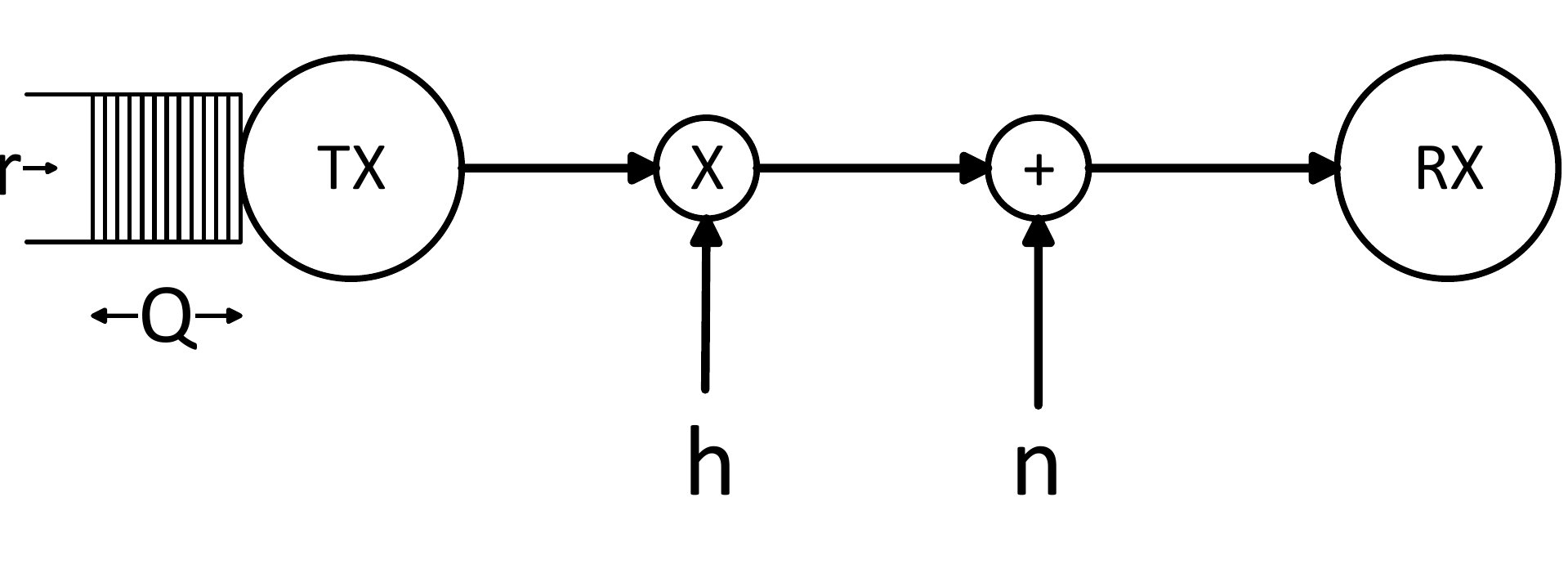}
	\centering
	\vspace{-0mm}
	\caption{System Model.}
	\vspace{-4mm}
	\label{f1}
\end{figure}


\subsection{Throughput of delay constrained networks}
It is considered that the data generated by random sources is stored as frames in the First in First out (FIFO) queue buffer before transmission as illustrated in Fig. \ref{f1}. Thus, the delay may occur in the transmission system because of the long waiting time of data in the buffer. Moreover, the delay overflow probability is given by \cite{120}
\begin{align}\label{eq3}
\text{Pr}\{D\geq d\}\approx\zeta e^{-\theta a(\theta) d},
\end{align}
where $D$ is the queueing delay, $d$ is the delay threshold, $\theta$ is the delay QoS constraint, $a(\theta)$ is the effective bandwidth and $\zeta$ is the probability of non-empty buffer. It is noted that larger value of $\theta$ means that stringent QoS constraint is imposed, while for lose QoS requirement the value $\theta$ is small \cite{shehab}.

In this paper, we discuss three types of Markov arrival sources namely Discrete-Time Markov source (DTMS), Fluid Markov source (FMS) and Markov Modulated Poisson source (MMPS). These Markov sources are focusing on two state ON and OFF model where the ON state corresponds to arrival with rate $\lambda$ bits/block and the OFF state refers to no arrival as illustrated in Fig. \ref{f2}. For these sources, effective bandwidth provides a mean to characterize the minimum constant service rates required to support the random arrival of data into the buffer constrained to some statistical QoS requirements, namely buffer violation probability in \eqref{eq3}. Let the time accumulated arrival process at instant $t$ be $A(t)=\sum_{k=1}^{t} a(k)$. Then the effective bandwidth is defined as \cite{108}
\begin{align}\label{eq4}
a(\theta)=\lim_{t\rightarrow\infty}{1\over \theta t}\log_{e}\mathbb{E}\{e^{\theta A(t)}\} \ ,
\end{align}

Effective capacity ($C_E$) is the dual concept of effective bandwidth, where it defines maximum constant arrival rate that a given time-varying service process can support in order to guarantee a statistical QoS requirement specified in the QoS exponent $\theta$. The effective capacity for a given QoS exponent is obtained from \cite{120}
\begin{align} \label{eq17}
C_{E}({ \gamma}, \theta)=-\lim_{t\rightarrow\infty}{1\over \theta t}\log_{e}\mathbb{E}\{e^{-\theta S[t]}\},
\end{align}
where $s[t]\triangleq \sum_{k=1}^{t} P[k]$ is the time accumulated service process and $\left\{P[k],k=1,2,\cdots\right\}$ shows the discrete time stationary and ergodic stochastic service process. So  $P[k]=r$, when it is ON state and otherwise 0 in OFF state.

 Performance analysis becomes quite important as well as challenging when the data arrival and channel response are random in the presence of QoS constraint. We are interested to find the maximum average arrival rate of Markovian sources that can support fading channel and satisfy the QoS requirement in \eqref{eq3}. The QoS requirement is satisfied when effective bandwidth of arrival process is equal to the effective capacity of service process \cite{main}, therefore
\begin{align}\label {eq11}
a(\theta)=C_{E}( \gamma,\mathrm{\theta}),
\end{align}
Then we can find maximum arrival rate that can support fixed rate transmissions at given SNR $\gamma$ and $\theta$. 

\section{Throughput of Markovian Source Models}
\subsection{Discrete - Time Markov Sources}
\begin{figure}[!t] 
	\centering \vspace{-0mm}
	\includegraphics[clip, trim= 2cm 13cm 1.5cm 7.5cm,width=0.65\columnwidth]{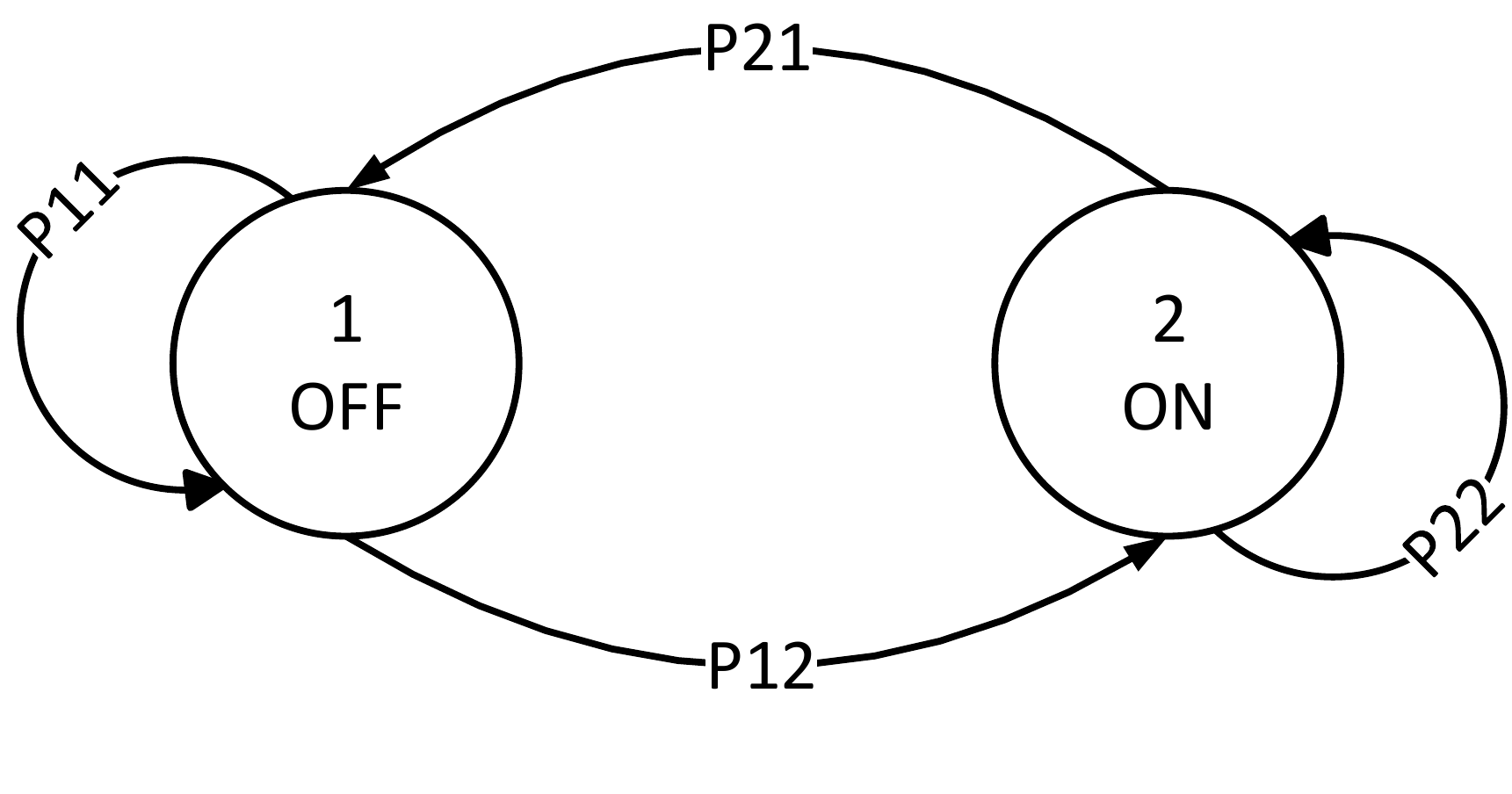}
	\centering \vspace{-0mm}
	\caption{ON-OFF state transition model.}
	\vspace{-3mm}
	\label{f2}
\end{figure}

In this model,  the arrival of data is discrete in time. We consider two-state simple (ON/OFF) model. When the state is ON, $\lambda$ bits arrive, while there are no arrivals in the OFF state. Effective bandwidth is expressed by \cite{111}
\begin{align}\label{eq5}
a(\theta)&={1\over \theta}\log_{e}\left({1\over 2}\left(p_{11}+p_{22}e^{\theta  \lambda}\right. \right. \notag \\
&\left.\left.+\sqrt{(p_{11}+p_{22}e^{\theta \lambda})^{2}-4(p_{11}+p_{22}-1)e^{\theta \lambda}}\right)\right), 
\end{align}
$p_{11}$ determines the probability of staying in OFF state, while $p_{22}$ identifies the probability of ON state as depicted in Fig. \ref{f2}. The transition probabilities from one state to another are denoted by $ p_{21}=1-p_{22} $ and $ p_{12}=1-p_{11}$.
$\mathrm{P_{ON}}$ is the  probability of ON state in the steady state regime, which is used to the calculate average arrival rate as 
\begin{align}\label {eq6}
\lambda_{\mathrm{avg}}= \lambda\cdot\mathrm{P_{ON}}=\lambda \ \frac{1-p_{11}}{2-p_{11}-p_{22}}\ ,
\end{align}
which is equal to the depature rate when the queue is in steady state \cite{110}.

We substitute the effective bandwidth expression of discrete time Markov source \eqref{eq5} in \eqref{eq11} and simplify as follows
\begin{align}\label{eq12}
(\rm p_{11}+p_{22}{e}^{\lambda \theta }-2{e}^{\theta C_{E}( \gamma,\theta)})^2&=(\rm p_{11}+p_{22}{e}^{\lambda\theta })^2 \\ \notag
&-4(\rm p_{11}+p_{22}-1){e}^{\lambda\theta }.
\end{align}
After solving \eqref{eq12} for $\lambda$, we obtain maximum ON state arrival rate as
\begin{align}\label{eq13}
\lambda^{*}(\theta)\!=\!\frac{1}{\theta}\log_{e}\!\left(\!\frac{{e}^{2\theta C_{E}( \gamma,\theta)}-p_{11}{e}^{\theta C_{E}( \gamma,\theta)}}{(1-p_{11}-p_{22})\!+\!p_{22}{e}^{\theta C_{E}( \gamma,\theta)}}\!\right).
\end{align}
Therefore, using \eqref{eq6} we expresses the maximum average arrival rate in terms of QoS exponent, effective capacity fading channel and state transition probabilities as
\begin{align}\label{eq14}
\lambda_{\mathrm{avg}}^*( \gamma,\theta)\!=\!\frac{\mathrm P_{\text{ON}}}{\theta}\log_{e}\!\left(\!\frac{{e}^{2\theta C_{E}( \gamma,\theta)}-p_{11}{e}^{\theta C_{E}( \gamma,\theta)}}{1\!-p_{11}\!-p_{22}+p_{22}{e}^{\theta C_{E}(\gamma,\theta)}}\!\right)\!.\!
\end{align} \vspace{1mm}

\subsection{Markov Fluid Sources}
In Fluid Markov Source model, the data arrival is continuous in time and the effective bandwidth is defined by \cite{main}  
\begin{align}\label{eq7}
a(\theta)\!=\!{1\over 2\theta}\left[\theta \lambda\!-\!(\!\alpha+\!\beta)\!+\!\sqrt{(\theta \lambda-(\alpha+\beta))^{2}+4\alpha\theta \lambda}\right]\!,
\end{align}
where $\alpha$ shows the transition rate from OFF state to ON state and $\beta$ is the transition rate from ON state to OFF state. Then, we attain the steady state probability of being ON as
\begin{align}\label {eq8}
\mathrm{P_{ON}}=\frac{\alpha}{\alpha + \beta }.
\end{align}
Furthermore, the average arrival rate is 
\begin{align}\label {eq9}
{\lambda_{\mathrm{avg}}}= \lambda \cdot \mathrm{P_{ON}}=\lambda \cdot \frac{\alpha}{\alpha + \beta }.
\end{align}
Similar to the previous source model, we identify the maximum average arrival rate of two state ON and OFF FMS source model as \vspace{-1mm} 
\begin{align}\label{eq15}
\lambda_{\mathrm{avg}}^*(\gamma,\theta)=\mathrm{P}_{\text{ON}}\frac{\theta C_{E}(\gamma,\theta)+\alpha+\beta}{\theta C_{E}(\gamma,\theta)+\alpha}C_{E}(\gamma,\theta).
\end{align}
\subsection{Markov Modulated Poisson Sources}
In this source model, the data arrival to the buffer is a Poisson process whose arrival intensity is controlled by continuous time Markov chain. Moreover, there is no arrival in OFF state (0 intensity), while in ON state $\lambda$ is the intensity of the Poisson arrival process. The effective bandwidth in this case is \cite{main}  \vspace{-2mm}
\begin{align}\label{eq10}
\begin{split}
a(\theta)&={1\over 2\theta}\left[\ (e^{\theta}-1) \lambda-(\alpha+\beta)\right]\\
&+{1\over 2\theta}\sqrt{(\ (e^{\theta}-1) \lambda-(\alpha+\beta))^{2}+4\alpha (e^{\theta}-1) \lambda}.
\end{split}
\end{align}
Similar as for previous sources models, we determine the maximum average arrival rate of two state ON and OFF MMPS source model as \vspace{-1mm}
\begin{align}\label{eq16}
\!\lambda_{\mathrm{avg}}^*(\gamma,\theta)\!=\! \mathrm P_{\text{ON}}\frac{\theta[\theta C_{E}(\gamma,\theta)  +  \alpha+ \!\beta]}{(e^\theta-1)\theta C_{E}( \gamma,\theta) +\alpha}C_{E}(\rm \gamma,\theta).
\end{align}


\section{Maximization of Effective Capacity}
Considering (\ref{eq5}) and noting that in our model $p_{11}+p_{22}=1$, we formulate the effective capacity for a given statistical QoS constraint $\theta$, as \cite{main2}
\begin{align} \label{eq20}
C_{E}({ \gamma}, \theta)&=-{1\over \theta }\log_{e}(p_{11}+p_{22}e^{-\theta r})\\
&=-{1\over \theta }\log_{e}(1-P\{z>\Psi\}(1-e^{-\theta r})),
\end{align}
where $\Psi=\frac{2^r-1}{\gamma}$. Under these assumption the channel  ON state probability is equivalent to  $ P \{z_i\! >\! \Psi \}$ and  OFF state  is $P \{z_i\! \leq\! \Psi\} $.
For Rayleigh fading, the pdf of $z$ is $p_z=e^{-z}$. Therefore \vspace{-2mm}
\begin{align}\label{eq22}
P\{z>\Psi\}&=(1-P\{z\leq\Psi\})\\
&=(1-\int_{0}^{\Psi}p_{z}(z)\ dz) \equiv e^{- \Psi}. 
\end{align}
Then plugging it into (19), we have
\begin{align}\label{eq24}
C_{E}({\rm \gamma},\!\theta)=-{1\over \theta }\log_{e}(1-e^{-\Psi}  \ (1-e^{-\theta r})).
\end{align}

We formulate the optimization problem to maximize $C_E$ subject to a non-negative transmission rate 
\begin{align}\label{eq25}
C_{E}^*({\rm \gamma}, \theta)&=\max_{r\geq 0}\left\{-{1\over \theta }\log_{e}(1-e^{-\Psi}  \ (1-e^{-\theta r}))\right\}.
\end{align}
Now, we obtain the first derivative of \eqref{eq25} as
\begin{align}\label{eq26}
\frac{\partial C_{E}({\rm \gamma}, \theta) }{\partial r}&= \frac{ \mathrm{\theta}\, \mathrm{e}^{-\Psi}\, \mathrm{e}^{-  r\, \mathrm{\theta}} + \frac{\log_{e}(2)\, 2^{r}\, \mathrm{e}^{-\Psi}\, \left(\mathrm{e}^{- r\, \mathrm{\theta}} - 1\right)}{{\gamma}}}{ \mathrm{\theta}\, \left(\mathrm{e}^{-\Psi}\, \left(\mathrm{e}^{-  r\, \mathrm{\theta}} - 1\right) + 1\right)}
\end{align}
and the second derivative is obtained in \eqref{eq27} on the top of the next page. which gives negative values only for rates around the upper contour of $C_E$. It means that $C_E$ is quasi concave in $r$. Therefore, the optimum value of $r$ is obtained by equating \eqref{eq26} to zero and solving for $r^*$ as follows
\begin{figure*}
	\begin{align}\label{eq27}
	\begin{split}
	\frac{\partial^2 EC }{\partial r^2}&=\frac{{\left(\mathrm{\theta}\, \mathrm{e}^{-{2^{r} - }\, {\gamma}}\, \mathrm{e}^{-\, r\, \mathrm{\theta}} + \frac{2^{r}\, \mathrm{e}^{-\Psi}\, \mathrm{\log_{e}}\!\left(2\right)\, \left(\mathrm{e}^{- r\, \mathrm{\theta}} - 1\right)}{ \gamma}\right)}^2}{ \mathrm{\theta}\, {\left(\mathrm{e}^{-\Psi}\, \left(\mathrm{e}^{ r\, \mathrm{\theta}} - 1\right) + 1\right)}^2}\\
	& - \frac{ {\mathrm{\theta}}^2\, \mathrm{e}^{-\Psi}\, \mathrm{e}^{ r\, \mathrm{\theta}} - \frac{2^{r}\, \mathrm{e}^{-\Psi}\, {\mathrm{\log_{e}}\!\left(2\right)}^2\, \left(\mathrm{e}^{ r\, \mathrm{\theta}} - 1\right)}{ \gamma} + \frac{2^{{2^r}}\, \mathrm{e}^{-\Psi}\, {\mathrm{\log_{e}}\!\left(2\right)}^2\, \left(\mathrm{e}^{ r\, \mathrm{\theta}} - 1\right)}{ {\gamma}^2} + \frac{2.\,2^{{r}}\ \mathrm{\theta}\, \mathrm{e}^{-\Psi}\, \mathrm{e}^{r\, \mathrm{\theta}}\, \mathrm{\log_{e}}\!\left(2\right)}{ \gamma}}{\mathrm{\theta}\, \left(\mathrm{e}^{-\Psi}\, \left(\mathrm{e}^{r\, \mathrm{\theta}} - 1\right) + 1\right)}.
	\hspace{-10mm}
	\end{split}
	\end{align}
	\hrule
\end{figure*}

\begin{align}\label{eq28}
&\frac{\gamma\, \theta\mathrm{e}^{-\Psi}\mathrm{e}^{-r\theta}+2^r\mathrm{e}^{-\Psi}\log_{e}(2)\left(\mathrm{e}^{- r\, \mathrm{\theta}} - 1\right)}{\gamma\, \mathrm{\theta}\left(\mathrm{e}^{-\Psi}\, \left(\mathrm{e}^{-  r\, \mathrm{\theta}} - 1\right) + 1\right)}=0.
\end{align}
After further simplification of \eqref{eq28}, we attain
\begin{align}
&\gamma\, \theta+2^r\log_{e}(2) = \frac{2^r\log_{e}(2)}{\mathrm{e}^{-r\theta}};
\end{align}
then, by taking logarithm in the both sides, we reach
\begin{align}\label{eq30}
\begin{split}
\begin{aligned}
&r^*=\frac{1}{\theta}\log_{e}\left(1+\frac{\gamma\, \theta}{2^{r^*}\log_{e}(2)} \right).
\end{aligned}
\end{split}
\end{align}
\eqref{eq30} gives a closed form solution of optimum transmission rate $r^*$. Plugging this value of $r^*$ in (23),  gives optimum effective capacity $C_E^*$, which in turn is used in \eqref{eq14}, \eqref{eq15} and \eqref{eq16} to obtain $\lambda^*_{\text{avg}}$ of different Markov sources.
\begin{figure}[h] 
	\vspace{-0mm}
	\centering
	\includegraphics[width=0.8\columnwidth]{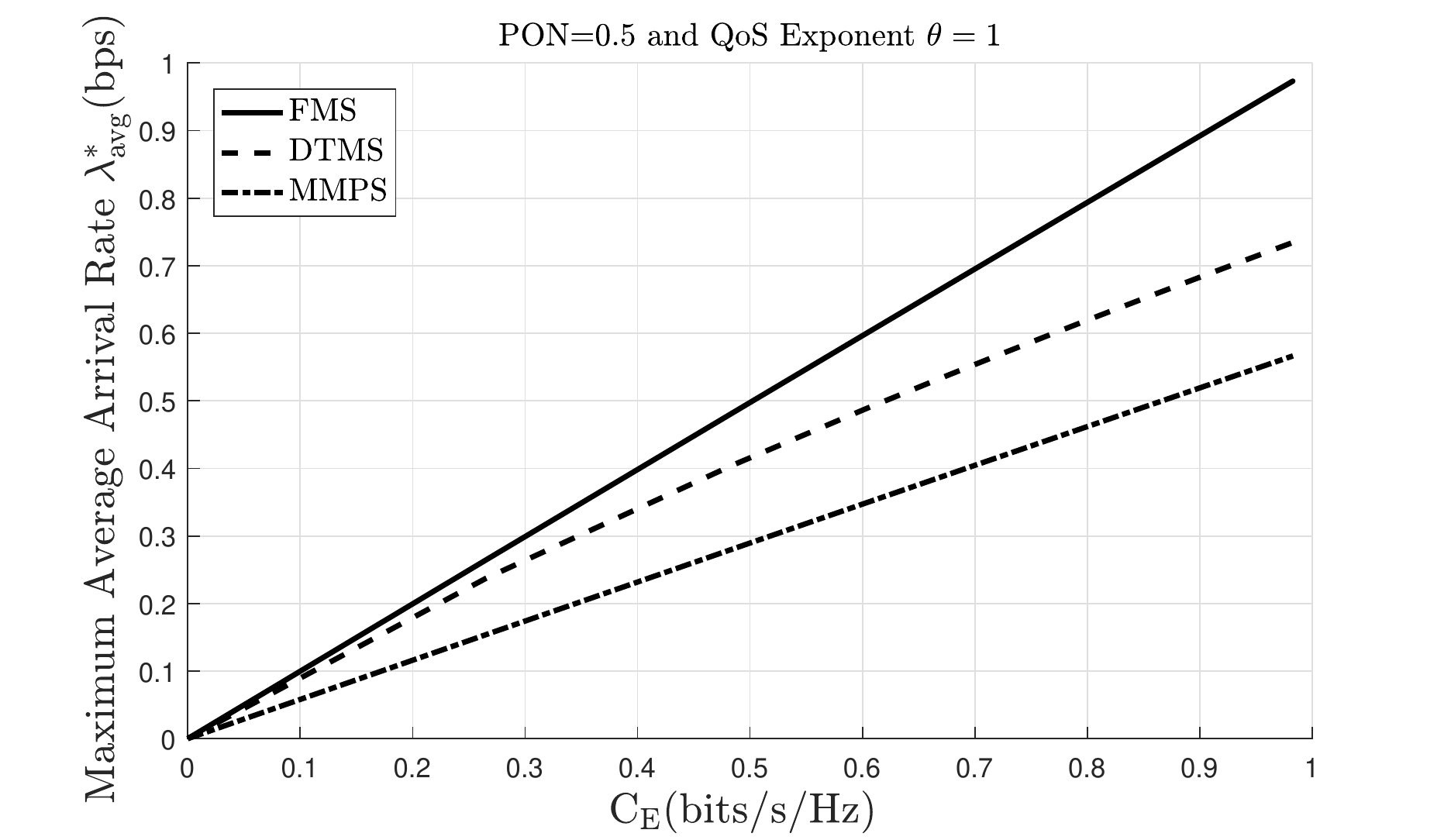}
	\caption{Maximum average arrival rate as a function of effective capacity for different Markovian source models.}
	\label{f3}
	\vspace{-1mm}
\end{figure}
\begin{figure}[h] 
	\centering
	\includegraphics[width=0.8\columnwidth]{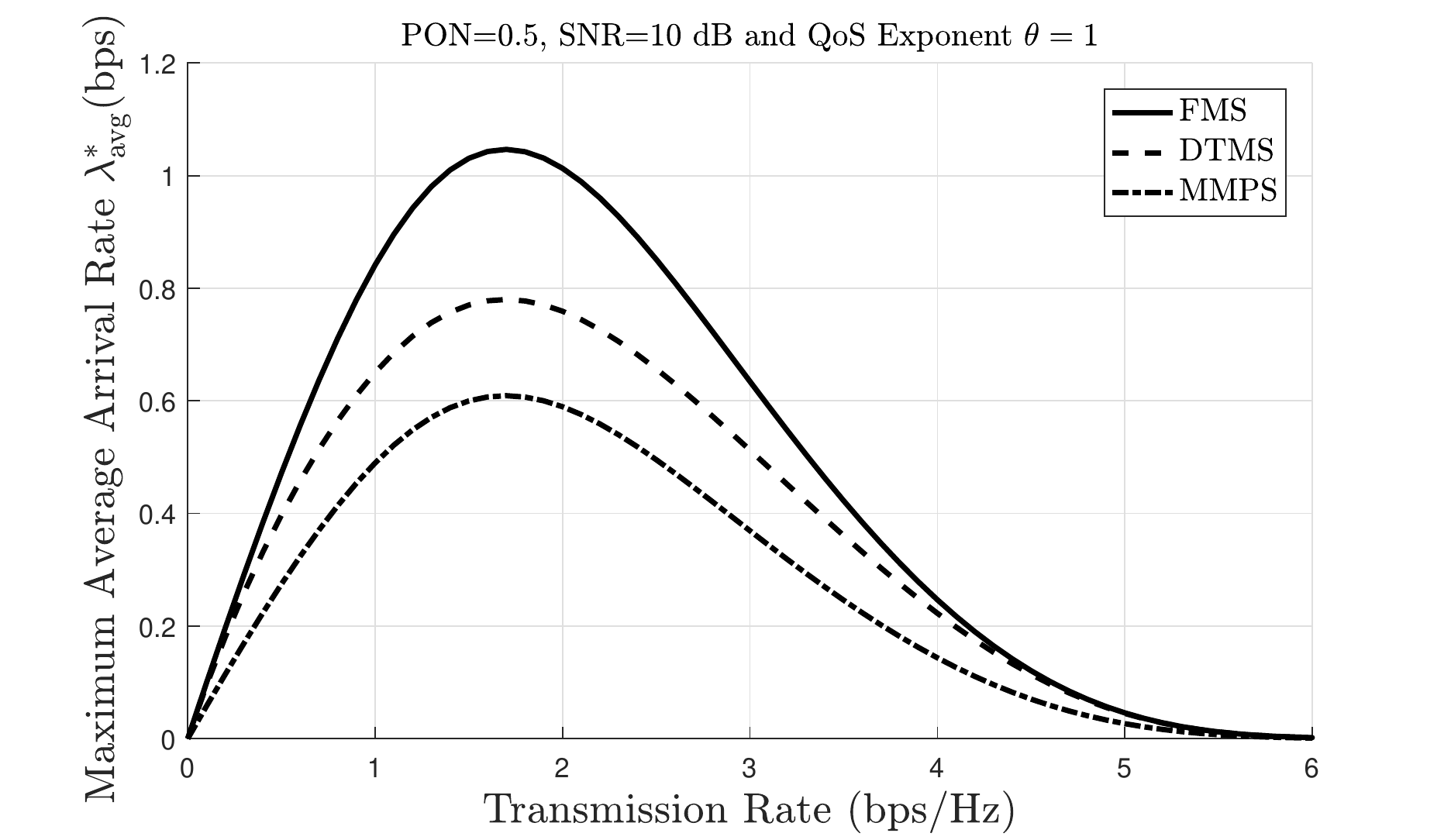}
	\vspace{-0mm}
	\caption{Maximum average arrival rate as a function of transmission rate for different Markovian source models.}
	\label{f4}
	\vspace{-1mm}
\end{figure}
\begin{figure}[h] 
	\centering
	\includegraphics[width=1\columnwidth]{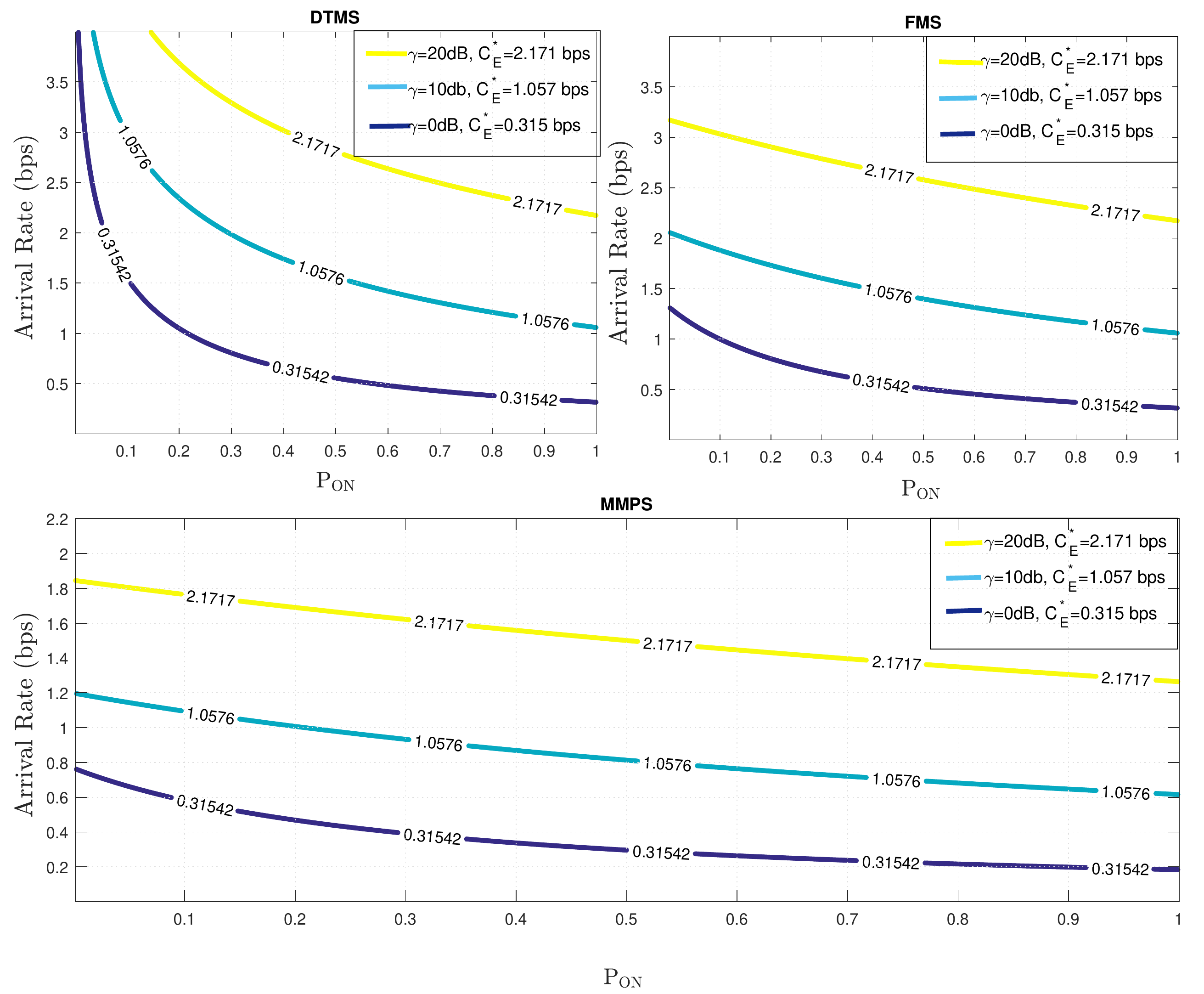}
	\caption{Arrival rate vs $\mathrm{P_{ON}}$ for different $\mathrm{C_E}$ of Markovian sources.}
	\label{f5}
	\vspace{-1mm}
\end{figure}
 
\section{results and discussion}
The efficient use of transmission rate boosts the performance of communication system. If we transmit the data with fixed rate, it may lead to waste of scarce resources. From Fig. \ref{f3}, it is clearly seen that $\lambda^*_{\text{avg}}$ of different Markov sources is monotonically increasing function of $C_E$ in a linear behaviour. So if we maximize, $C_E^*$ we also maximize $\lambda^*_{\text{avg}}$. Thus, there a unique rate which can maximize $C_E$ and consequently maximize the $\lambda^*_{\text{avg}}$. Fig. \ref{f4}, shows $\lambda^*_{\text{avg}}$ as a function of transmission rate with $\gamma=10$ dB for different Markovian  source models. As already proven, it is obvious that $C_E$ and consequently $\lambda^*_{\text{avg}}$ are quasi concave functions of the rate because the upper contour set is convex.

Furthermore, we numerically investigate the throughput of effective rate transmission model with ON-OFF Markov arrival processes in the presence of QoS requirements as we assume Rayleigh fading channel. In Fig. \ref{f5}, we plot the arrival rate levels vs. $\mathrm{P_{ON}}$ curves that required to support given  $C_E^*$ for different value of  $\gamma \in  \{0, 10, 20\}  \mathrm{dB}$, when QoS exponent $\theta=1$. It is  noticed that when the source is always ON i.e, $\mathrm{P_{ON}=1}$, then the maximum arrival rate equal to ${C_E^*}$. We also illustrate that as $\mathrm{P_{ON}}$ diminishes, arrival rate in ON state needs to increase with a certain level in order to keep average arrival non-decreasing. However, with same departure rate it is difficult to keep throughput non-decreasing, when QoS constraints are imposed. Therefore, higher arrival rate is required to achieve ${C_E^*}$ when the source becomes bursty (i.e $\mathrm{P_{ON}<1}$). We also observe that MMPS  has more tolerance as compare to DTMS and FMS under the arrival of bursty/random source. For instance with $\gamma$=10 dB and $\mathrm{P_{ON}=0.2}$, DTMS provide $C_E^*$ 1.057 bps with arrival rate of 2.34 bps while, MMPS supports an arrival rate of 1.02 bps which is closer to $C_E^*$. Hence DTMS and FMS are more affected by the random/bursty sources as they require a significant adaption of the arrival rate to guarantee QoS when the source is more bursty. So far, we have observed mainly the impact of burstiness/randomness to the effective transmission rate framework.

\begin{figure}[h!] 
	\centering
	\includegraphics[width=1\columnwidth]{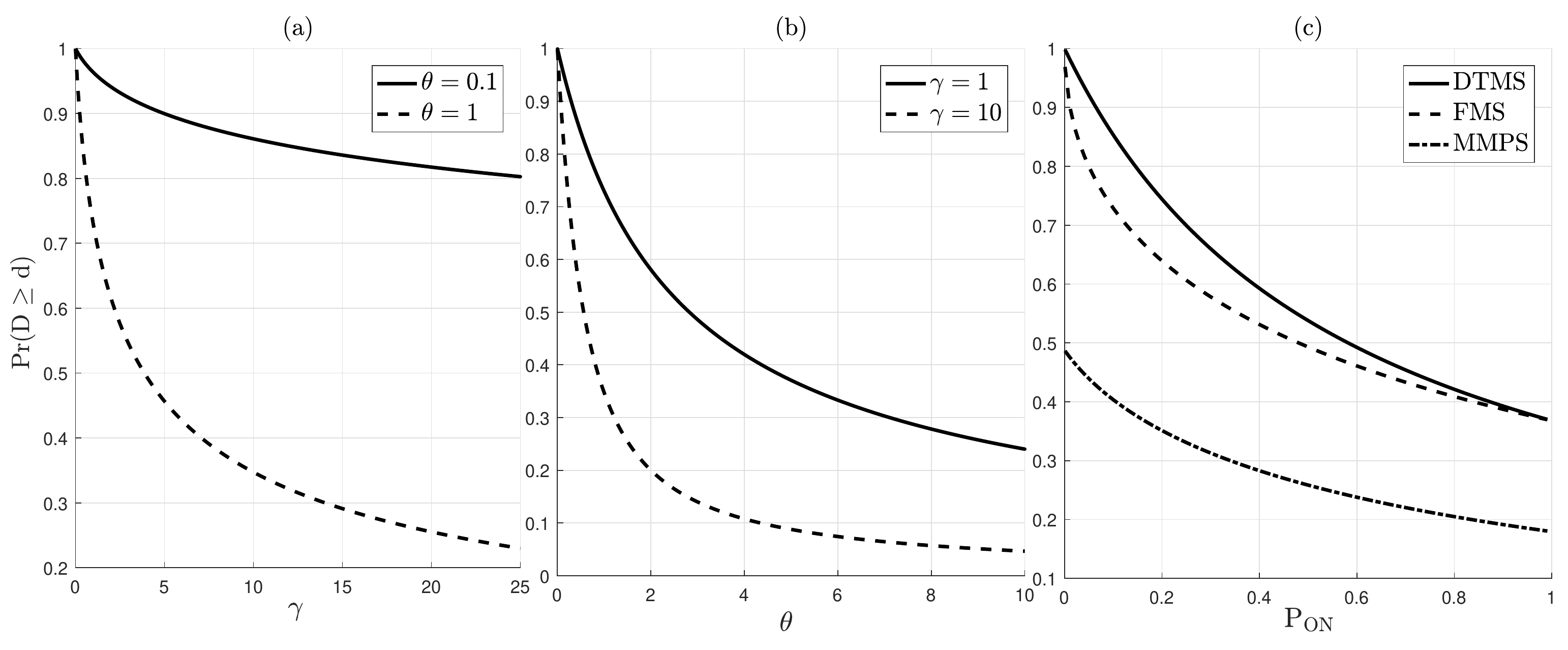}
	\vspace{-2mm}
	\caption{Delay violation probability vs (a) $\gamma$, (b) $\theta$ and (c) $\mathrm {P_{ON}}$.}
	\label{f6}
	\vspace{-2mm}
\end{figure}
In Fig. \ref{f6}, we are interested to evaluate effects of latency, where delay violation probability is showed vs $\gamma$, $\theta$ and $\mathrm {P_{ON}}$. We notice that delay violation probability decreases logarithmically with the increase in $\gamma$, $\theta$  and $\mathrm {P_{ON}}$. In Fig. \ref{f6}(a),
as SNR increases, more channel capacity is available for data to fulfill latency requirement of the system. Therefore probability of delay violation is reduced. Fig. \ref{f6}(b) shows that as QoS requirement increases, the delay violation probability is reduced. In Fig. \ref{f6}(c), delay violation probability is plotted as a function of $\mathrm {P_{ON}}$ for fixed arrival rate of 1 bps and fixed target reliability and latency outage probability. Burstiness is measured from the average arrival of data in ON state. It is observed that more bursty sources degrade the maximum average arrival rate, which directly rises delay violation probability. It is clearly noticed that MMPS sources tolerate low delay violation probability in the presence of random/bursty sources as compared DTMS and FMS.

\vspace{-1mm}

\section{Conclusion} \label{conclusions}
In this work, we formulated effective transmission rate model to achieve adequate reliability and latency requirement in single point-to-point machine type devices. We incorporated Markov source models to investigate their performance over sources arrival traffic and Rayleigh fading channel. The source, buffer, and channel characteristics have major impact on the performance of model when certain QoS constraints are imposed. Moreover, we introduced a throughput metric that captures the behaviour of sources burstiness and channel condition in the design of effective transmission link when certain level of reliability and latency is required. The results showed that MMPS models is more robust to bursty sources than FMS while DTMS is the least robust. Moreover, increased source burstiness and stringent QoS requirement all need an increase in SNR gain to fulfill reliable communication. \vspace{-1mm}
\section*{Acknowledgments}
This work is partially supported by Academy of Finland 6Genesis Flagship (Grant no. 318927), Aka Project SAFE (Grant no. 303532), and by Finnish Funding Agency for Technology and Innovation (Tekes), Bittium Wireless, Keysight Technologies Finland, Kyynel, MediaTek Wireless, and Nokia Solutions and Networks.

 \bibliographystyle{IEEEtran}
 \bibliography{fahad}

\end{document}